\documentclass[%
 reprint,
superscriptaddress,
 amsmath,amssymb,
 pra,
]{revtex4-1}

\usepackage{graphicx}
\usepackage{dcolumn}
\usepackage{bm}
\usepackage{braket}
\usepackage{color}
\usepackage{subcaption}
\usepackage[font=footnotesize,labelfont=bf]{caption}
\usepackage[font=footnotesize,labelfont=bf]{subcaption}
\usepackage{hyperref}

\begin{document}

\preprint{APS/123-QED}

\title{Preparing the spin-singlet state of a spinor gas in an optical cavity}

\author{Stuart J. Masson}\email{smas176@aucklanduni.ac.nz} 
\affiliation{Dodd-Walls Centre for Photonic and Quantum Technologies, New Zealand}
\affiliation{Department of Physics, University of Auckland, Auckland, New Zealand}

\author{Scott Parkins}\email{s.parkins@auckland.ac.nz}
\affiliation{Dodd-Walls Centre for Photonic and Quantum Technologies, New Zealand}
\affiliation{Department of Physics, University of Auckland, Auckland, New Zealand}

\date{\today}

\begin{abstract}
We propose a method to prepare the spin singlet in an ensemble of integer-spin atoms confined within a high-finesse optical cavity. Using a cavity-assisted Raman transition to produce an effective Tavis-Cummings model, we show that a high fidelity spin singlet can be produced probabilistically, although with low efficiency, heralded by the {\it absence} of photons escaping the cavity. In a different limit, a similar configuration of laser and cavity fields can be used to engineer a model that emulates spinor collisional dynamics. Borrowing from techniques used in spinor Bose-Einstein condensates, we show that adiabatic transformation of the system Hamiltonian (via a time-dependent, effective quadratic Zeeman shift) can be used to produce a low fidelity spin singlet. Then, by following this method with the aforementioned heralding technique, we show that it is possible to prepare the singlet state with both high fidelity and good efficiency for a large ensemble.
\end{abstract}

\maketitle

\section{Introduction}

Spinor Bose gases - in particular, ensembles of ultracold Bose atoms with internal spin degrees of freedom - offer a remarkable platform for the study of quantum fluids and phenomena such as quantum phase transitions and superfluidity \cite{Kawaguchi12,StamperKurn13}. One reason for this is the rich variety of collision-induced, spin-mixing dynamics that are possible in spinor Bose-Einstein condensates (BECs), together with the exotic quantum states that may result from these dynamics. Of particular interest in this context, driven in part by potential application to quantum-enhanced metrology, have been highly entangled (e.g., spin squeezed) states \cite{Pu00,Duan02,Mustecaplioglu02,Sau10,Zhang13}, and remarkable experimental progress has been made in this field of research \cite{Lucke11,Gross11,Hamley12,Hoang13,Peise15,Hoang16NatComm,Linnemann16,Kruse16,Hoang16PNAS,Luo17,Zou18}.

One system that has been of interest is a small, tightly confined BEC of spin-1 atoms in the presence of a magnetic field. In the single mode approximation, the interplay between collisions and the quadratic Zeeman shift is modeled by the Hamiltonian
\begin{equation}
\hat{H} = \frac{\Lambda}{N} \mathbf{\hat{S}}^2 + q \hat{N}_0 , \label{hamiltonianeqn}
\end{equation}
where $\mathbf{\hat{S}} = (\hat{S}_x,\hat{S}_y,\hat{S}_z)$ and $\hat{S}_i$ are collective angular momentum operators for the ensemble of $N$ spin-1 atoms, $\hat{N}_0$ is the population operator for the $m=0$ state, and $\Lambda$ and $q$ characterize the interaction strength and quadratic Zeeman shift, respectively. 

Despite its apparent simplicity, the Hamiltonian (\ref{hamiltonianeqn}) admits a variety of ground states dependent on the signs and relative magnitudes of $\Lambda$ and $q$. One of those ground states is the macroscopic spin singlet. This is a state of $N$ atoms where the \emph{collective} angular momentum is zero. This state features strong entanglement and is given by the superposition
\begin{align}
\ket{S=0} &= \sum\limits_{j=0}^{N/2} c_{j} \ket{j,N-2j,j}, 
\end{align}
using the notation $\ket{n_{-1},n_0,n_{+1}}$ for $n_i$ atoms in the magnetic state $m=i$ and with coefficients given by
\begin{align}
c_0 = \frac{1}{\sqrt{N+1}}, &~~~~ c_j = - \sqrt{\frac{N-2j+2}{N-2j+1}} c_{j-1}.
\end{align}
This state features genuine multipartite entanglement of the entire ensemble \cite{Vitagliano11,Vitagliano14}. This entanglement is useful for quantum metrology \cite{UrizarLanz13} and could be of use in fields including quantum memory \cite{Lidar98} and quantum information processing \cite{Cabello02,Bartlett03}.

Preparing the spin singlet of a BEC as the ground state of Hamiltonian (\ref{hamiltonianeqn}) is, in practice, a very challenging prospect, given the extremely small energy scales involved \cite{StamperKurn13}. However, other methods have been proposed to produce the spin-singlet state. One such method involves a sequence of quantum non-demolition measurements using pulses of light to probabilistically prepare a highly entangled macroscopic spin singlet \cite{Behbood13,Toth10,Hauke13}. 

In this work, we propose a method to produce the macroscopic spin singlet via interactions mediated by cavity-assisted Raman transitions. The imbalanced Dicke model for spin-1 atoms \cite{Zhiqiang17,Masson17}, or more simply a Tavis-Cummings model \cite{Masson18PhotonCounting}, can probabilistically produce the spin singlet heralded by the {\it absence} of measured photons in the cavity output. This protocol works with an efficiency equal to the overlap between the initial state and the spin singlet, which, however, is $1/(N+1)$. We thus propose the use of quasi-adiabatic sweep techniques - an established method for these systems in spinor BECs - to enhance that initial overlap. We then show that a protocol of a quasi-adiabatic sweep followed by the probabilistic distillation of the state using a Tavis-Cummings model offers a method to produce a singlet state with very high fidelity and a reasonably high efficiency.


\section{Setup and model}

\begin{figure}[t]
\includegraphics[width=0.35\textwidth]{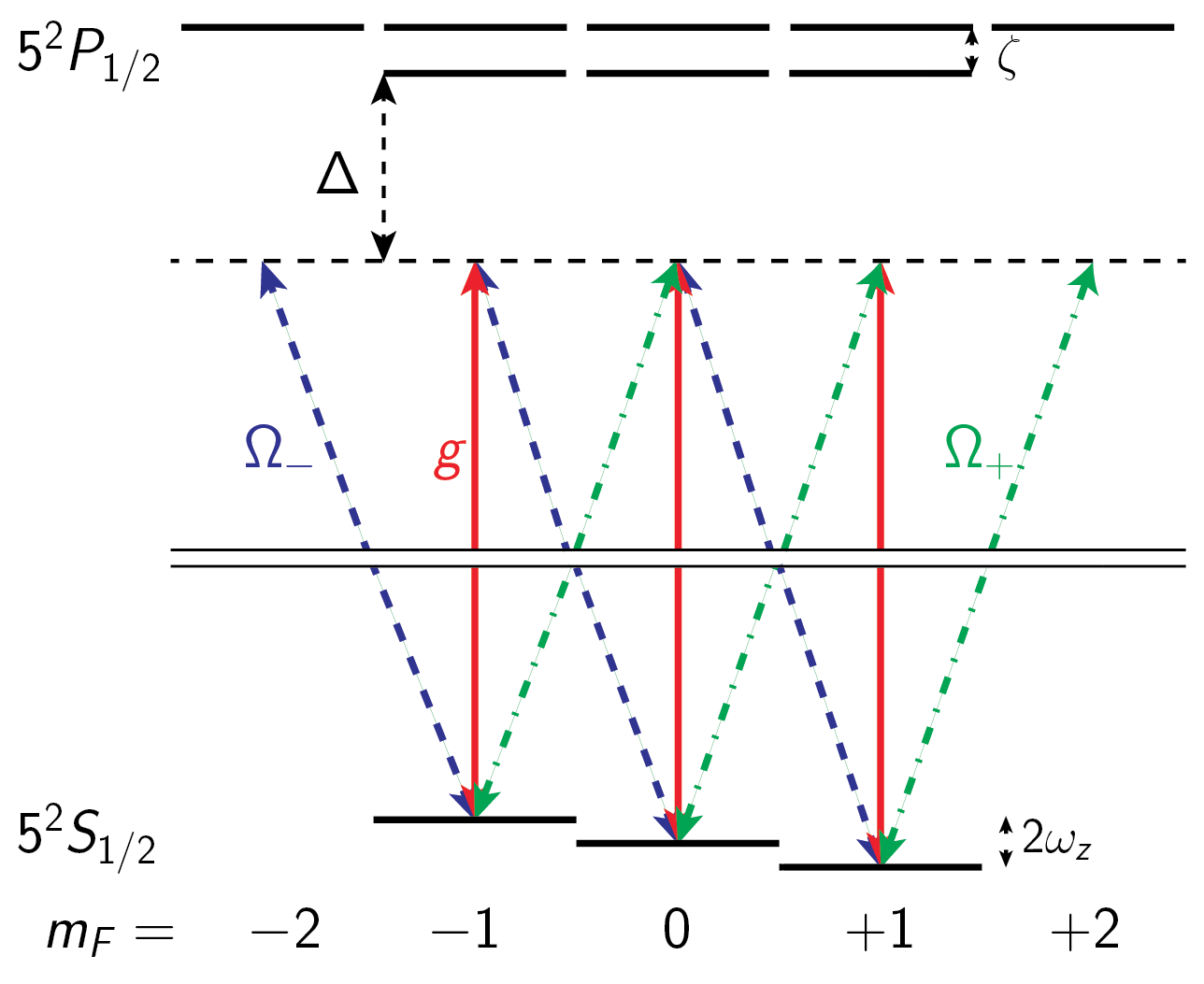}
\caption{Implementation of an effective spin-1 Dicke model using the $F=1$ ground state of $^{87}$Rb. Interactions are mediated by detuned Raman transitions on the $D_1$ line composed of a cavity mode (red) and $\sigma_-$- (blue) and $\sigma_+$- (green) polarized lasers.\label{leveldiagram}}
\end{figure}

We consider an ensemble of $^{87}$Rb atoms held tightly within an optical cavity. We assume that the ensemble is sufficiently dilute that we can ignore any direct atom-atom interactions, and instead engineer effective interactions via cavity-assisted Raman transitions \cite{Dimer07,Morrison08PRL,Morrison08PRA,Zhiqiang17,Masson17,Zhiqiang18,Masson18PhotonCounting}, as illustrated in Fig.~\ref{leveldiagram}. Here, as demonstrated recently in \cite{Zhiqiang17}, we consider transitions within the complete $F=1$ hyperfine ground state. In the limit that the detunings of the Raman transitions are much larger than the energy separations of the excited state hyperfine levels, this gives an effective open Dicke model for an ensemble of spin-1 particles. In that case, the evolution of the density operator can be described by the master equation \cite{Zhiqiang17,Masson17,Masson18PhotonCounting}
\begin{equation}
\dot{\rho} = -i[\hat{H},\rho] + \kappa \mathcal{D}[\hat{a}] \rho \label{dickemodelmasterequation},
\end{equation}
where $\hat{a}$ is the cavity mode annihilation operator, $\mathcal{D}[\hat{a}]\rho = 2\hat{a}\rho\hat{a}^\dagger - \rho \hat{a}^\dagger \hat{a} - \hat{a}^\dagger\hat{a}\rho$, and
\begin{eqnarray}\label{dickemodelH}
\hat{H} &&= \omega \hat{a}^\dagger \hat{a} + \omega_0 \hat{S}_z \nonumber
\\
&& ~+ \frac{\lambda_-}{\sqrt{2N}} (\hat{a}\hat{S}_+ + \hat{a}^\dagger \hat{S}_-) + \frac{\lambda_+}{\sqrt{2N}} (\hat{a}\hat{S}_- + \hat{a}^\dagger\hat{S}_+).
\end{eqnarray}
The coefficients of the various terms are determined by light shifts and Raman transition rates; in particular,
\begin{align}
\omega &= \omega_c - \frac{\omega_- + \omega_+}{2} + \frac{Ng^2}{3\Delta} , \\
\omega_0 &= \omega_z - \frac{\omega_- - \omega_+}{2} + \frac{\Omega_+^2 - \Omega_-^2}{24\Delta} ,\\
\lambda_\pm &= \frac{\sqrt{N}g\Omega_\pm}{12\Delta}.
\end{align}
Here, $\omega_c$ is the frequency of the cavity mode, $\omega_\pm (\Omega_\pm)$ are the bare frequencies (Rabi frequencies) of the $\sigma_\pm$-polarized laser fields, $\omega_z$ is the linear Zeeman splitting of the $F=1$ levels, $g$ is the single-atom-cavity coupling strength (for the $^{87}$Rb $D_2$ line cycling transition), $\kappa$ is the cavity field decay rate, and $\Delta$ is the detuning of the fields from the atomic resonance. Note that we assume that the atoms all couple to the cavity mode with the same strength $g$, which can in practice be achieved by confining the atoms tightly at cavity mode antinodes in an optical lattice potential (which may be created via another, far-detuned cavity mode at twice the wavelength of the other mode \cite{Zhiqiang17,Zhiqiang18}).

By turning off the $\sigma_+$-polarized laser and choosing $\omega =\omega_0$, the above model reduces to the resonant Tavis-Cummings model for spin-1 atoms. We will consider this model in Section IV in the context of a dissipative, probabilistic scheme for preparing the spin-singlet state.

Alternatively, we can consider the above model in a dispersive limit, such that $\omega\gg\omega_0, \lambda_\pm$, in which case the cavity mode can be adiabatically eliminated. Considering again the case where $\Omega_+=0$, the model can then be reduced to \cite{Masson17,Davis18,Marino18}
\begin{equation}
\dot{\rho} = -i[\hat{H},\rho] + \frac{\Gamma}{N} \mathcal{D}[\hat{S}_-]\rho,\label{equationdickemodelspinor}
\end{equation}
with
\begin{equation}\label{eq:H}
\hat{H} = \omega_0' \hat{S}_z + \frac{\Lambda}{N} (\hat{S}_x^2 + \hat{S}_y^2) ,
\end{equation}
and in terms of the Dicke model parameters given above, we define new parameters
\begin{equation}
\omega_0' = \omega_0 + \frac{\Lambda}{N}, ~~~ \Lambda = - \frac{\omega\lambda^2}{2(\omega^2 + \kappa^2)}, ~~~ \Gamma = -\frac{\kappa}{\omega}\Lambda.
\end{equation}
By manipulating the microscopic parameters, it is possible to (at least approximately) set $\omega_0' = 0$, but given that we consider a system initiated with all atoms in the $m=0$ level (i.e., with $\braket{\hat{S}_z} = 0$), and that $\hat{S}_z$ is conserved by the Hamiltonian evolution, this term does not impact the evolution in the absence of photon detections, which we show is the heralding condition for the production of the singlet. Hence, we see that the cavity-mediated, coherent spin interactions described by (\ref{eq:H}) can emulate the collisional interactions of BECs in the single mode approximation.

Further to this, an artificial quadratic Zeeman shift can, for example, be produced by a weak, auxiliary $\pi$-polarized laser field acting near the $F'=1$ excited manifold. This shift is considered to be time dependent, and could be adjusted either by moving it closer or further from resonance, or by adjusting the power of the weak field. This leaves us with a Hamiltonian of the form
\begin{equation}
\hat{H} = \frac{\Lambda}{N} (\hat{S}_x^2 + \hat{S}_y^2) - q(t) \hat{N}_0.\label{hamiltoniantimedep}
\end{equation}

\section{Entanglement criteria}

Due to the additional degrees of freedom for particles with spin $>1/2$, entanglement in spinor particles can be quantified by a range of different inequalities \cite{Vitagliano11,Vitagliano14}. From \cite{Vitagliano14}, we have that an ensemble of spin-1 particles with no entanglement satisfies
\begin{equation}
(\Delta \hat{S}_x)^2 + (\Delta \hat{S}_y)^2 + (\Delta \hat{S}_z)^2 \geq N.
\end{equation}
Therefore, if an ensemble breaks the less strict bound
\begin{equation}
\braket{\hat{S}_x^2} +\braket{ \hat{S}_y^2} + \braket{\hat{S}_z^2} \geq N \rightarrow \braket{\hat{\mathbf{S}}^2} \geq N,\label{vitaglianocriterion}
\end{equation}
then that ensemble is entangled. This measure thus has two purposes: it shows us how the spin length decreases as well as acting as an entanglement witness. Furthermore, $\braket{\hat{\mathbf{S}}^2}$ gives a bound on the maximum number of atoms that are not entangled \cite{Toth10}.

\section{Dissipative evolution\label{dissipation}}

We consider the master equation (\ref{dickemodelmasterequation}) with $\omega = \omega_0 = 0$. An initial state with all the atoms in the $m=0$ state can be decomposed into a superposition of Dicke states $\ket{S,0}$ with different spin lengths as (for even $N$ \footnote{Our scheme also works for the $F=2$ hyperfine ground state, i.e., an ensemble of spin-2 atoms, where the initial state $\ket{m=0}^{\otimes N}$ has a finite overlap with the spin-singlet state regardless of whether the number of atoms is even or odd. Spin-1 atoms are considered here for simplicity.})
\begin{align}\label{eq:psi0}
\ket{n=0} \otimes \ket{0,N,0} &= \ket{n=0} \otimes \sum\limits_{k=0}^{N/2} d_k \ket{2k,0} ,
\end{align}
where the distribution $\{ d_k\}$ is strongly peaked around $k\simeq\sqrt{N}$ \cite{Masson18PhotonCounting}.
Excitations are produced in the cavity in conjunction with a spin ladder operator, i.e., via terms in the Hamiltonian of the form $\hat{a}^\dagger \hat{S}_\pm$, and that ladder operator does not operate on the spin singlet (i.e., $\hat{S}_\pm \ket{0,0}=0$). This means that entanglement is generated between the cavity mode and the atoms: an empty cavity with the spin singlet, and non-zero photon numbers with the states of non-zero spin length occurring in (\ref{eq:psi0}). Any photon emitted from the cavity must therefore collapse the state into a superposition of spin states not containing the spin singlet, while a null measurement will project the atomic state into the spin singlet. Monitoring the cavity output thus gives some probability of projecting the state into the spin singlet; the probability is simply the overlap of the initial state with the spin singlet, which is $1/(N+1)$ for the state (\ref{eq:psi0}).

\begin{figure}[t]
\begin{subfigure}{0.45\textwidth}
\includegraphics[width=\textwidth]{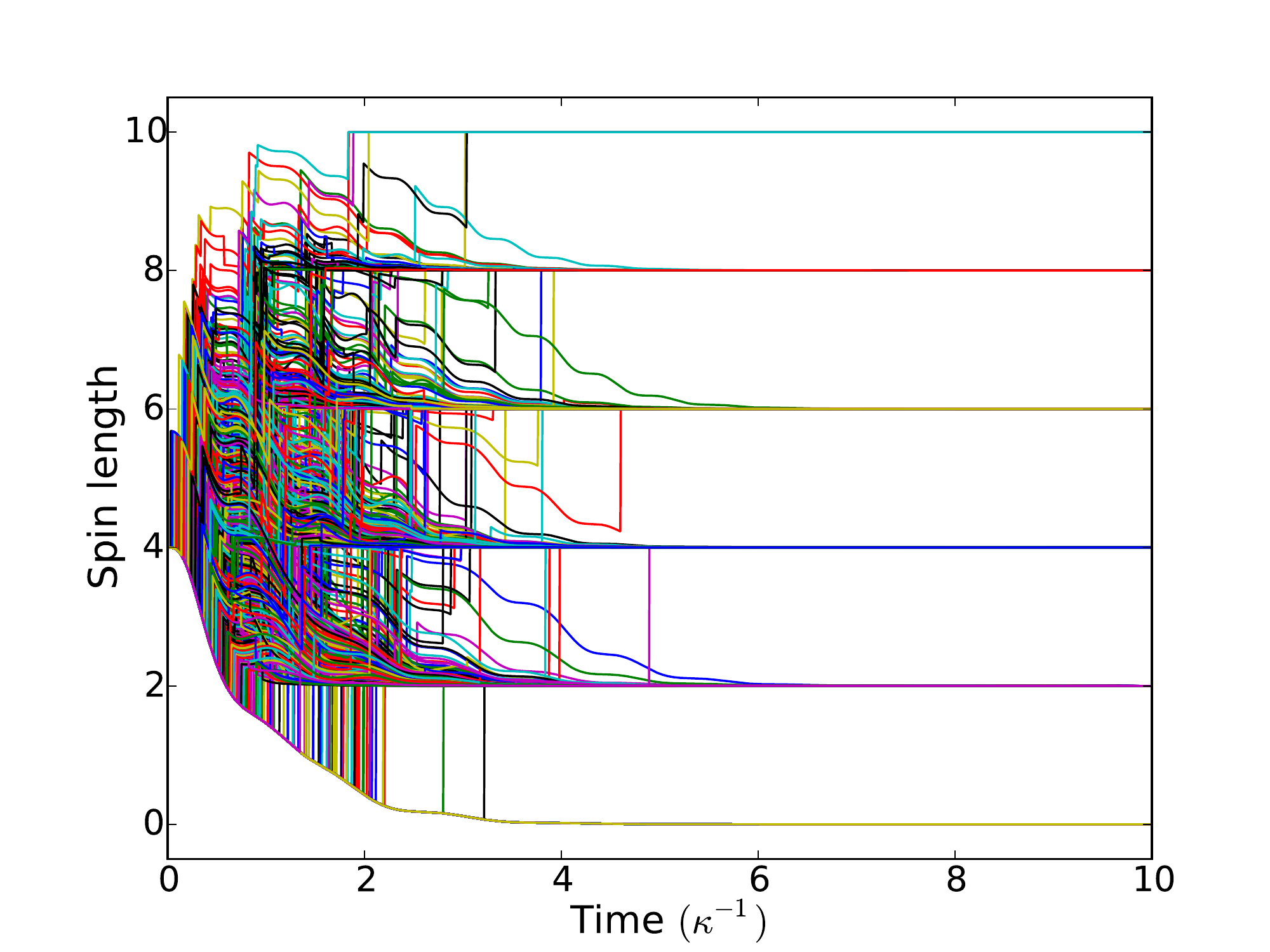}
\caption{}
\end{subfigure}

\begin{subfigure}{0.45\textwidth}
\includegraphics[width=\textwidth]{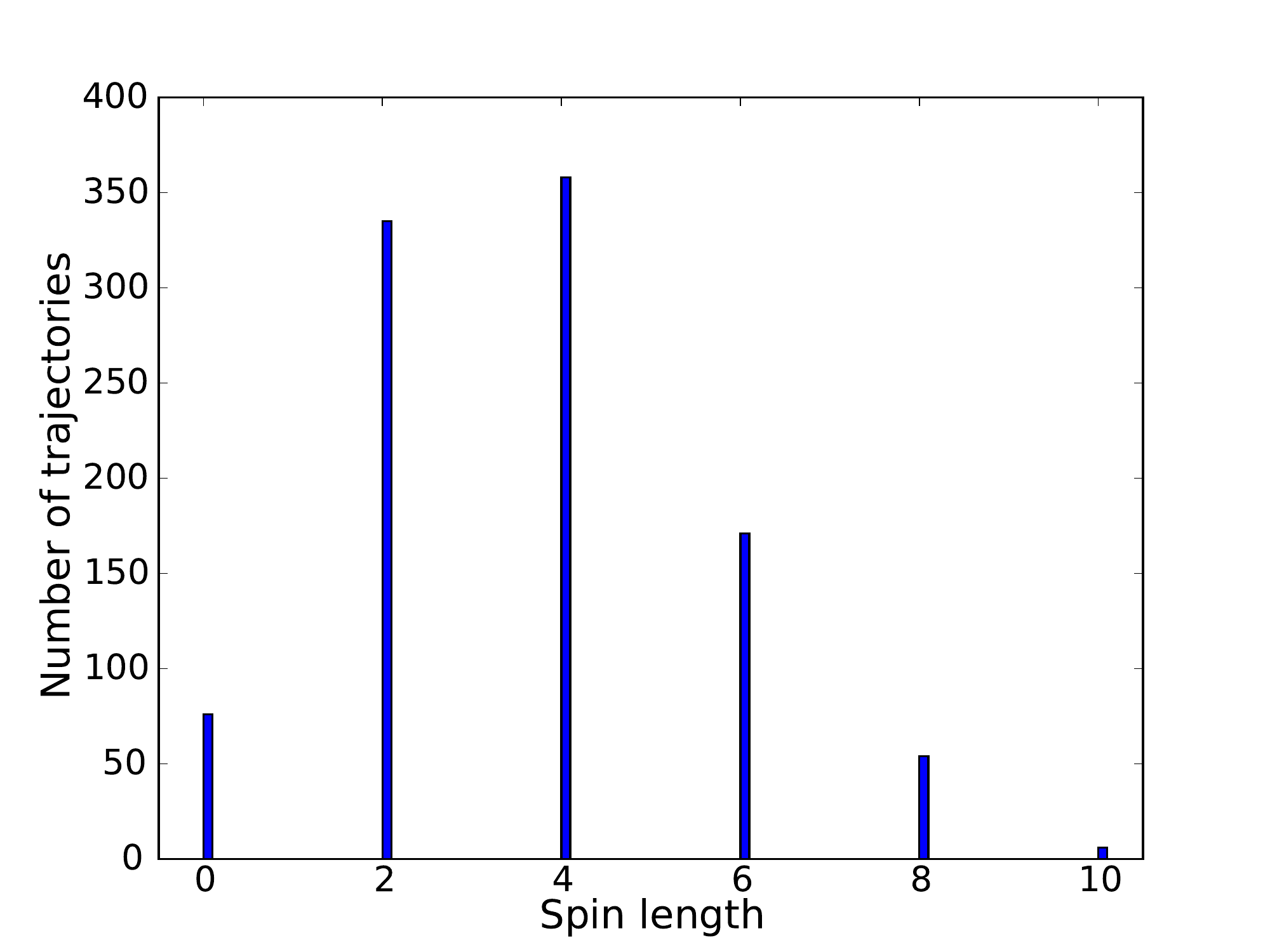}
\caption{}
\end{subfigure}
\caption{Spin length, found by calculating $\braket{\hat{\mathbf{S}}^2} = S(S+1)$ and solving for $S$, (a) over time and (b) at $\kappa t = 10$, in 1000  quantum Monte-Carlo trajectories for $N=10$ atoms under master equation (\ref{dickemodelmasterequation}) with $\lambda_- = 6\kappa$, and $\omega = \omega_0 = \lambda_+ = 0$.\label{dissipationplot}}
\end{figure}

If we turn off one of the lasers; in particular, if we set $\lambda_+ = 0$, then an initial atomic state $\ket{2k,0}$ will evolve, subject to Eqs.~(\ref{dickemodelmasterequation}) and (\ref{dickemodelH}) (i.e., the damped Tavis-Cummings model, when $\lambda_+ = 0$), to a steady state $\ket{2k,- 2k}$, with the emission of $2k$ photons from the cavity \cite{Masson18PhotonCounting}. An ideal way in which to study this behavior is to use the method of Monte Carlo wave function simulations (or quantum trajectories) \cite{Carmichael1993}, and in Fig.~\ref{dissipationplot} we present results of this approach applied to the model of (\ref{dickemodelmasterequation}) and (\ref{dickemodelH}), where each ``jump'' in the simulations corresponds to the emission of a photon from the cavity. Fig.~\ref{dissipationplot}(a) shows that, after a few cavity lifetimes, the state is always cleanly projected to a definite spin length. After a time $\kappa t = 10$ with $N=10$ atoms, the overlap with the spin singlet is either (essentially) one or zero, depending on the output photon record (Fig.~\ref{dissipationplot}(b)).

The projection, for the parameters of Fig.~\ref{dissipationplot}, occurs on a timescale on the order of the photon lifetime in the cavity, $1/\kappa$. This is because the production of photons in the cavity due to coupling to the atoms happens on a much faster timescale than cavity loss, and the rate of photon emission follows the rate of the slower process. For this to hold for arbitrary atom number we require that photon production happens for \emph{all} states at a rate faster than $\kappa$. The slowest state to produce photons is $\ket{S=2,0}$, and so we require $\lambda_- \sqrt{3/N} \gg \kappa$. If that is not the case, the rate will instead be governed by the intracavity photon production rate, and so the singlet would be heralded by the absence of photon detections over some timescale more than a few cavity lifetimes.

With a more realistic photon detection scheme that has an efficiency $\eta$, the absence of emitted photons from the Tavis-Cummings system would project to a mixed state (a normalizing factor is omitted),
\begin{align}
\rho = &|d_0|^2 \ket{0,0}\bra{0,0} \notag \\&+ \sum\limits_{k=1}^{N/2} (1-\eta)^{2k} |d_k|^2 \ket{2k,-2k} \bra{2k,-2k}.
\end{align}
For reasonably high efficiencies (in particular high enough such that $(1-\eta)^{2}|d_2|^2 \ll |d_0|^2$), this should still be dominated by the spin singlet. Alternatively, the Dicke model ($\lambda_+=\lambda_-$), which produces a continuous stream of photons, could be used. This would allow for very high fidelity singlet production even without perfect detection efficiency. It should be noted that very high effective atom-cavity coupling, $\lambda_\pm$, or very long measurement times would be necessary to project out the lower spin state contributions.

The probability of this procedure working is simply the initial overlap with the singlet state, $1/(N+1)$. For small numbers of atoms this process offers an efficient method to prepare a highly entangled state. However, for 1000 atoms, the maximum efficiency would be 0.1$\%$. Other experimental considerations, such as photon detection efficiency, would further reduce the overall efficiency.

This may be mitigated somewhat by multiple runs with a single atomic ensemble. Using a feedback system conditioned on a photon detection (or some threshold depending on dark count rates), switching back on the repumping system to reinitialize the state into the state $\ket{0,N,0}$ would reintroduce an overlap with the spin singlet. Since the time required for each run is relatively short, this should allow for multiple runs over the course of the lifetime of the ensemble.


\section{Quasi-adiabatic methods in spinor BECs}

If we want to improve the efficiency of our procedure, we need to improve the overlap between the initial state and the spin singlet. To do so, we can make use of the various ground states admitted by the spin-collisional Hamiltonian (\ref{hamiltonianeqn}). In the limit where the quadratic Zeeman shift dominates over the collisional interaction, i.e., $|q| \gg |\Lambda|$, and $q$ is negative, the ground state of the system has all atoms in the $m=0$ state, i.e., $\ket{0,N,0}$, which is of course readily prepared via suitable optical pumping. If, instead, $q$ is positive, then the ground state is degenerate between all states for which there are no atoms in the $m=0$ state. Depending on the spread through the $\hat{S}_z$ states that satisfy this, the degeneracy includes entangled states, such as the twin Fock state $\ket{N/2,0,N/2}$, as well as completely classical states such as $\ket{N,0,0}$ or coherent combinations. 
 
 In the other limit, where instead the collisional interaction dominates over the quadratic Zeeman shift, i.e., $|\Lambda| \gg |q|$, there are again two ground states. With ferromagnetic interactions ($\Lambda < 0$, as for $^{87}$Rb) the ground state is degenerate for all states with a maximum spin length $N$. As above, this can range from a highly entangled Dicke state to a completely classical state, depending on how $\hat{S}_z$ is constrained. For anti-ferromagnetic interactions ($\Lambda > 0$, as for $^{23}$Na), there is only one ground state -- the spin singlet. 

The existence of these different ground states allows, in theory, for preparation of entangled states by adiabatic passage. For example, an ensemble can be prepared in the (unentangled) state $\ket{0,N,0}$, which is the ground state for large, negative $q$. An adiabatic sweep from there to $q=0$ produces, depending on the sign of $\Lambda$, either the spin singlet \cite{Zhang13,Hoang16PNAS} or, due to the conservation of $\hat{S}_z$, the Dicke state $\ket{S = N, S_z = 0}$ \cite{Zou18}. Alternatively, an adiabatic sweep through to large, positive $q$ would prepare the twin Fock state $\ket{N/2,0,N/2}$ \cite{Zhang13,Luo17}.

However, a key issue with performing this experimentally is maintaining adiabaticity throughout the sweep. At the phase transition points, the energy gap between the ground  and first excited states becomes extremely small (zero in the limit $N\rightarrow\infty$). Since an adiabatic sweep requires parameters to change on a time scale very slow compared to the inverse of the energy gap, the sweep has to be extremely slow through the transition. Hence, a true adiabatic sweep typically faces severe challenges associated with achievable experimental run times. 

Hoang et al. \cite{Hoang16PNAS} used an ensemble of 40000 atoms and a sweep time of 35~s (the minimum time for true adiabaticity). After such a long time, less than 25\% of the original BEC remained trapped. Due to the noise that atom loss induces in the magnetization, no measurable entanglement was left in the ensemble. The alternate approach is to ramp faster than adiabatic, introducing a small amount of energy to the system. This moves the state out of the true ground state, but into states near the ground state that still feature high entanglement. Luo et al. \cite{Luo17} instead ramped in 3~s to high negative $q$ attempting to produce the twin Fock state, finding just $4\%$ of the atoms remained in the $m=0$ state. Such a quasi-adiabatic sweep has also been shown to produce metrologically useful entanglement in a BEC \cite{Zou18}.

\section{Hamiltonian evolution\label{hamiltonian}}

\begin{figure}[b!]
\begin{subfigure}{0.495\textwidth}
\includegraphics[width=\textwidth]{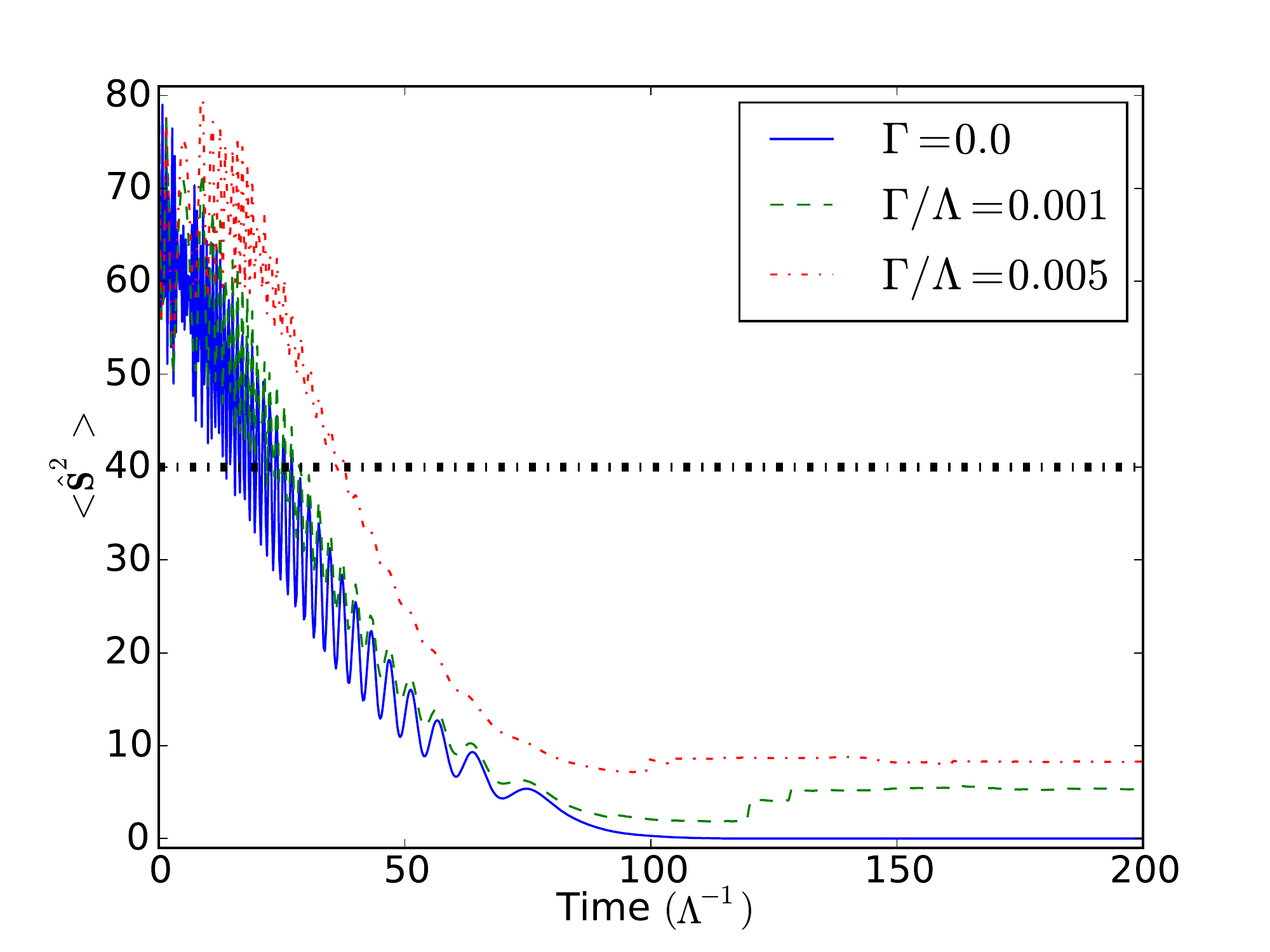}
\caption{\label{exposweepa}}
\end{subfigure}

\begin{subfigure}{0.495\textwidth}
\includegraphics[width=\textwidth]{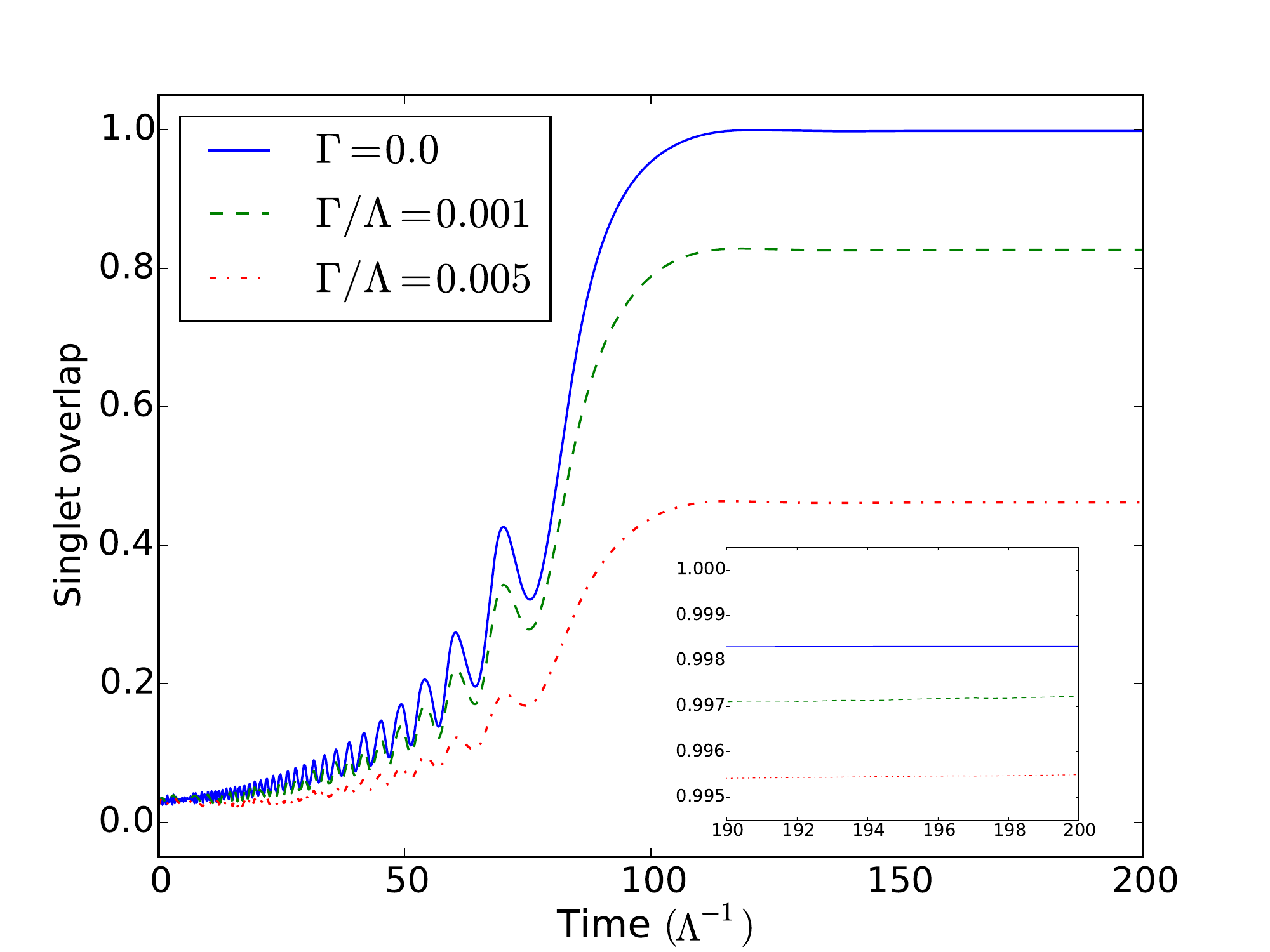}
\caption{\label{exposweepb}}
\end{subfigure}
\caption{Simulations for $N=40$ atoms. (a) Spin length squared and (b) overlap with the spin-singlet state for a sweep $q(t)/\Lambda = 7.0\mathrm{e}^{-0.08\Lambda t}$ with and without losses. Results with losses, i.e. non-zero $\Gamma$, are plotted as an ensemble average of 1000 trajectories. Inset zooms in on the overlap at later times conditioned on zero jumps.\label{exposweep}}
\end{figure}

We now consider evolution with the engineered master equation (\ref{equationdickemodelspinor}) with the Hamiltonian (\ref{hamiltoniantimedep}). Here, the intention is to emulate results from BECs by quasi-adiabatically sweeping the quadratic Zeeman shift, characterized by $q$, from some large value to essentially zero. To truly maximize the overlap, we should optimize the sweep such that it is as close to adiabatic as possible in the given timeframe \cite{Hoang16PNAS,Luo17,Zou18}. However, in the interest of simplicity, we have instead considered just sweeps that are straightforward in form, but that still greatly enhance the overlap with the spin singlet. We find that an exponential decay, $q(t)/\Lambda = q_0 \mathrm{e}^{-\xi \Lambda t}$, produces a much higher fidelity spin singlet compared to linear-in-time or reciprocal-in-time decays.

In Fig.~\ref{exposweep}, we use an average over an ensemble of quantum trajectories to approximate the master equation solution and show that the sweep greatly increases the overlap with the spin singlet, and, concomitantly, greatly reduces the spin length. We also see that the non-adiabaticity adds energy to the system, resulting in large oscillations of the spin length during the sweep. These oscillations, and, in particular, their phase as they are ``frozen out'' by the quadratic Zeeman shift settling at zero, add considerable noise to the resultant spin length and overlap.

\begin{figure}[b!]
\begin{subfigure}{0.46\textwidth}
\includegraphics[width=\textwidth]{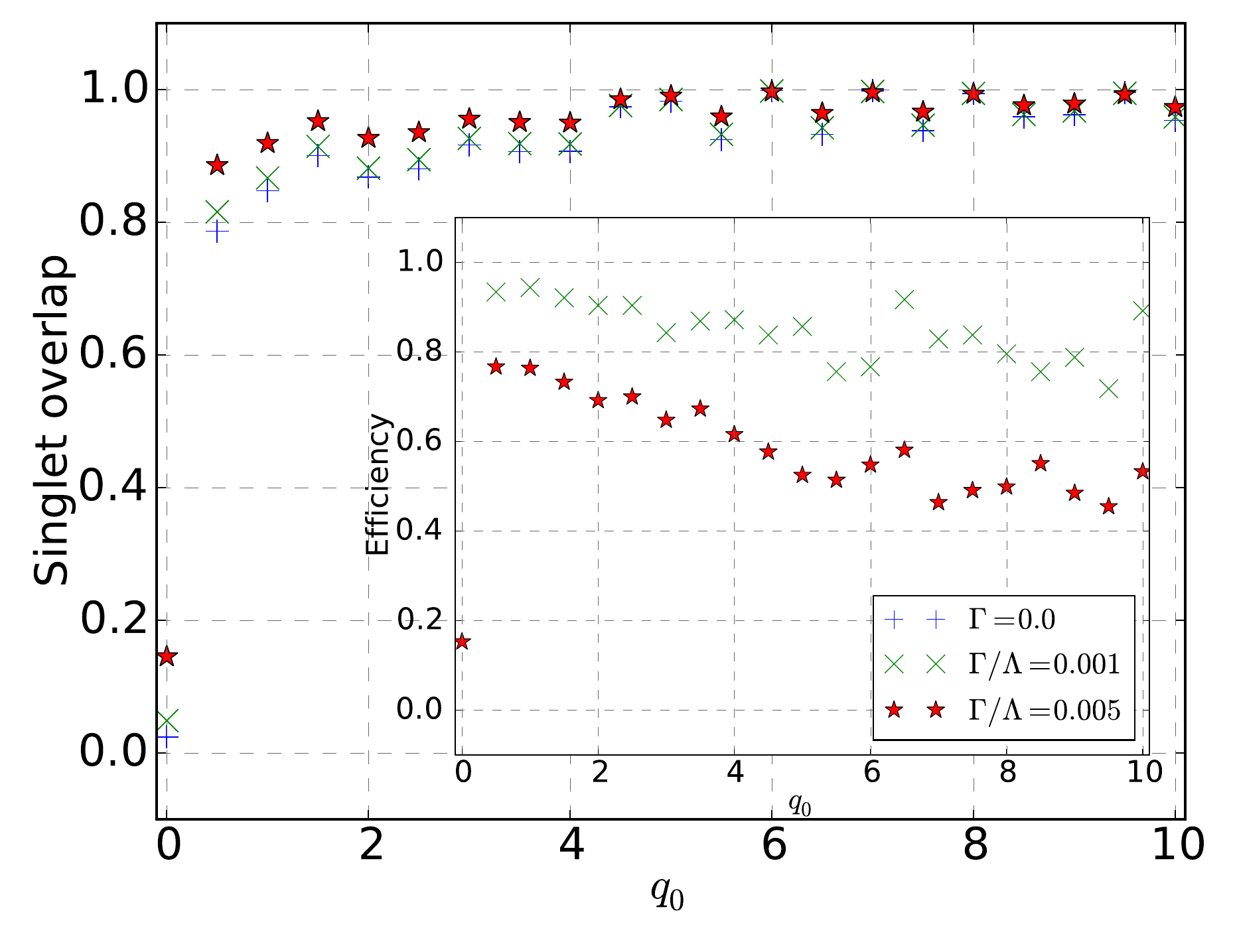}
\caption{\label{exposweepc}}
\end{subfigure}

\vspace{4mm}

\begin{subfigure}{0.46\textwidth}
\includegraphics[width=\textwidth]{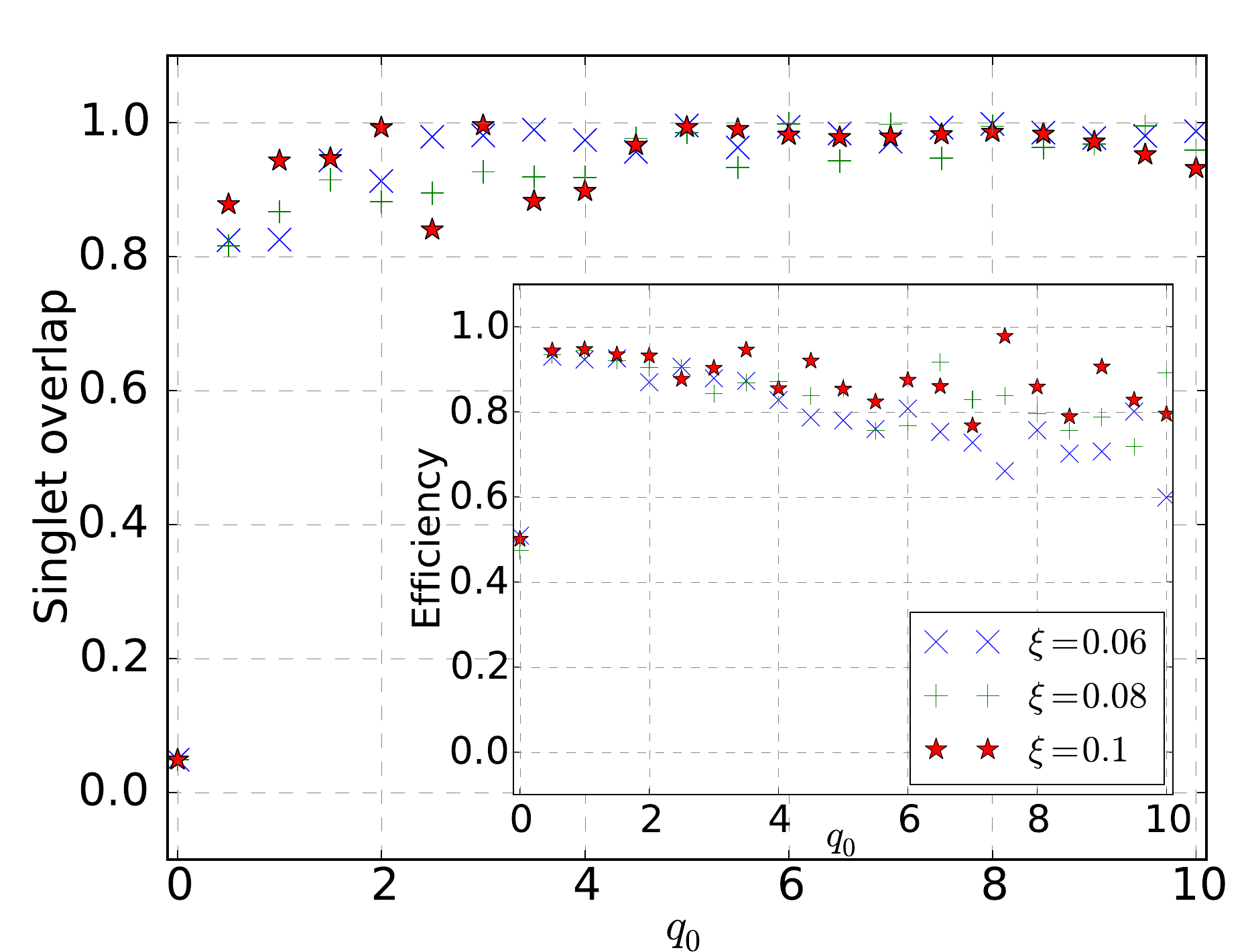}
\caption{\label{exposweepe}}
\end{subfigure}
\caption{Simulations for $N=40$ atoms. (a,b) Final overlap with the spin singlet with zero emitted photons and (insets) post-selection efficiency (i.e., no photons emitted) given (a) $q(t)/\Lambda = q_0 \mathrm{e}^{-0.08 \Lambda t}$, or (b) $q(t)/\Lambda = q_0 \mathrm{e}^{-\xi \Lambda t}$ with $\Gamma / \Lambda = 0.001$. Post-selection efficiencies are the percentage of 1000 trajectories that would produce zero photons. \label{exposweep2}}
\end{figure}

Using quantum trajectories, we find that the results can be split into two categories: with jumps and without jumps. With even a single jump, corresponding to the emission of a cavity photon, the overlap with the spin singlet becomes zero. This is because the spin singlet is a dark state to the jump operator, $\hat{S}_-$. Trajectories without a jump exhibit high overlap with the singlet state; in fact, for some sweep parameters, the overlap is higher than without losses. This is because the null measurement back-action increases the overlap with the dark state of the system: the spin singlet.

In line with the work in Section \ref{dissipation}, there is the possibility of post-selecting runs with a null photon output measurement, i.e., no jumps. As the spin length increases, so does the number of photons such a state produces. In particular, due to the spread of spin length, certain states in the initial superposition can release huge numbers of photons, greatly increasing $\braket{\hat{\mathbf{S}}^2}$ for that trajectory. For example, with $\Gamma/\Lambda = 0.001$ in Fig.~\ref{exposweep}, we find that 829 of the 1000 trajectories have no jumps, and so have very high overlap with the spin singlet. Of the other 171 trajectories, there are 139 with $\braket{\hat{\mathbf{S}}^2} \approx 6$, i.e., $S\approx2$, commensurate with the production of one or two photons. All bar two trajectories have $\braket{\hat{\mathbf{S}}^2} < 40$, with the remaining two having a final value $\braket{\hat{\mathbf{S}}^2} \sim 1600$. These states therefore each contribute a significant portion of the ensemble-averaged result, but, even with a realistic single photon detector, could easily be discarded by post-selection, since they emit $\sim 40$ photons to attain such a spin length.

Considering Fig.~\ref{exposweep}, we obtain, with $\Gamma = 0$, a state with all 40 atoms entangled, while with $\Gamma>0$ (i.e., finite cavity decay), we have more than 30 atoms entangled according to the criterion (\ref{vitaglianocriterion})  introduced earlier in the paper. Notably, the entanglement bound is strongly violated even for the master equation result, i.e., with no post-selection, and even with the highest considered cavity-mediated, damping $\Gamma / \Lambda = 0.005$.

Choosing the parameters for the sweep involves striking a balance between a slow ramp, finishing with the quadratic Zeeman shift close to zero, and limiting the cavity-mediated losses. However, clear dependences are not obvious due to the oscillations introduced by the non-adiabatic sweep. Instead, wide regions of sweep parameters offer reasonable results. We can see in Fig.~\ref{exposweepc} that it is difficult to pinpoint an ideal value of the initial quadratic Zeeman shift, $q_0$, with a general trend that $q_0 \gtrsim 4\Lambda$ produces higher overlap. However, higher $q_0$ reduces the probability of a trajectory without photon loss. Fig.~\ref{exposweepe} shows that the rate of the exponential decay, $\xi$, is also difficult to choose so as to optimize the overlap, though, broadly speaking, higher $\xi$ gives a higher post-selection efficiency.

We note that, of course, optimising the quasi-adiabatic sweep in more complex ways should enhance the overlap of post-selected trajectories. However, such optimized sweeps might not necessarily be ideal for the probability of zero photons being produced during the sweep and so that would need to be taken into consideration for the optimisation process.


\section{Combination}

We now show that for large atom numbers, the dissipative scheme can offer reasonable efficiency if it is preceded by the sort of quasi-adiabatic singlet preparation discussed above. The total efficiency of the scheme is now the post-selection efficiency of the sweep (i.e., the probability of no photons being emitted during the sweep) multiplied by the efficiency of the dissipative scheme, which is the overlap at the end of the sweep,
\begin{equation}
p = p_s |\braket{\psi | S=0}|^2 .
\end{equation}

\begin{figure}[t!]
\begin{subfigure}{0.235\textwidth}
\includegraphics[width=\textwidth]{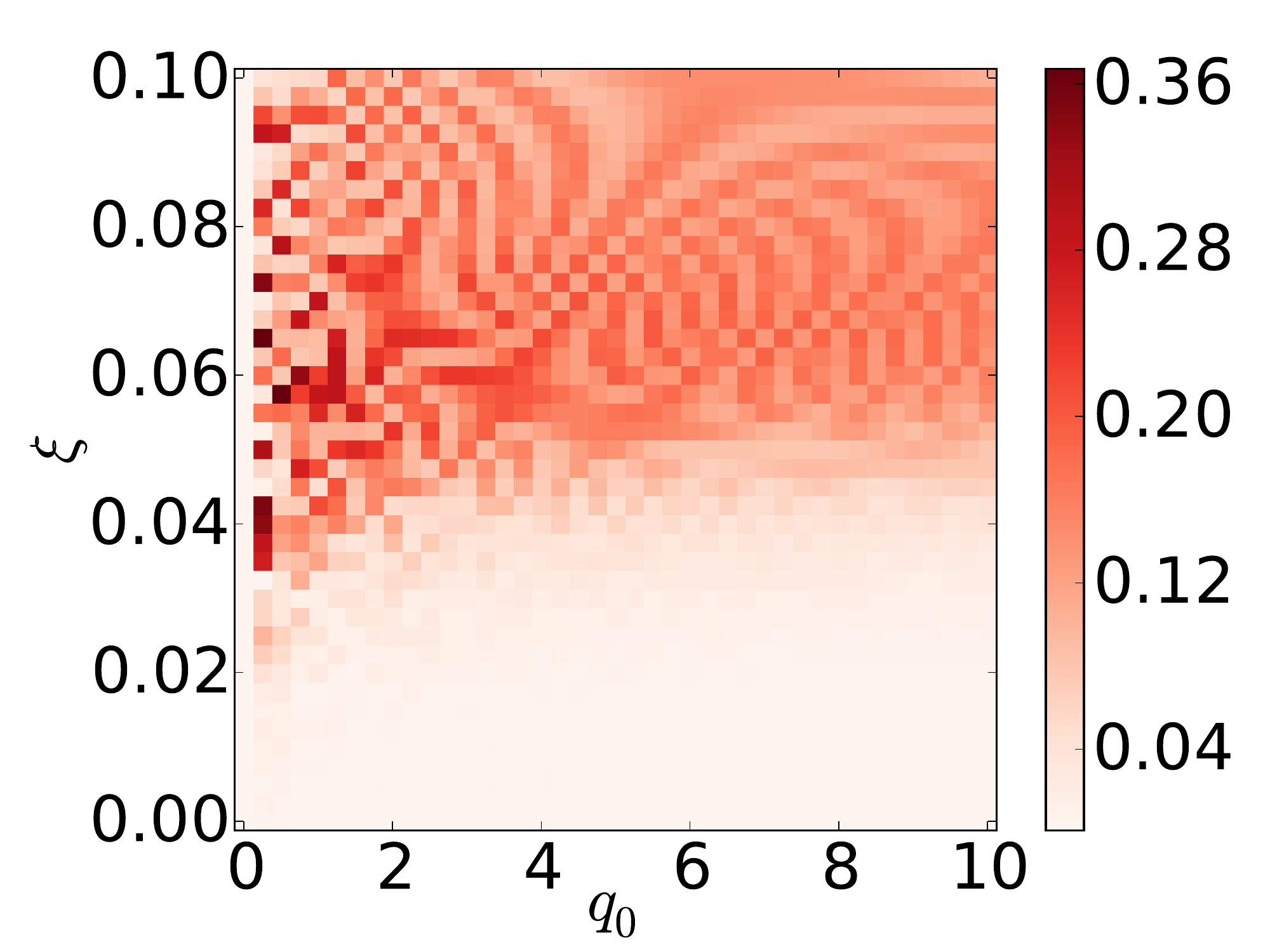}
\caption{}
\end{subfigure}
\begin{subfigure}{0.235\textwidth}
\includegraphics[width=\textwidth]{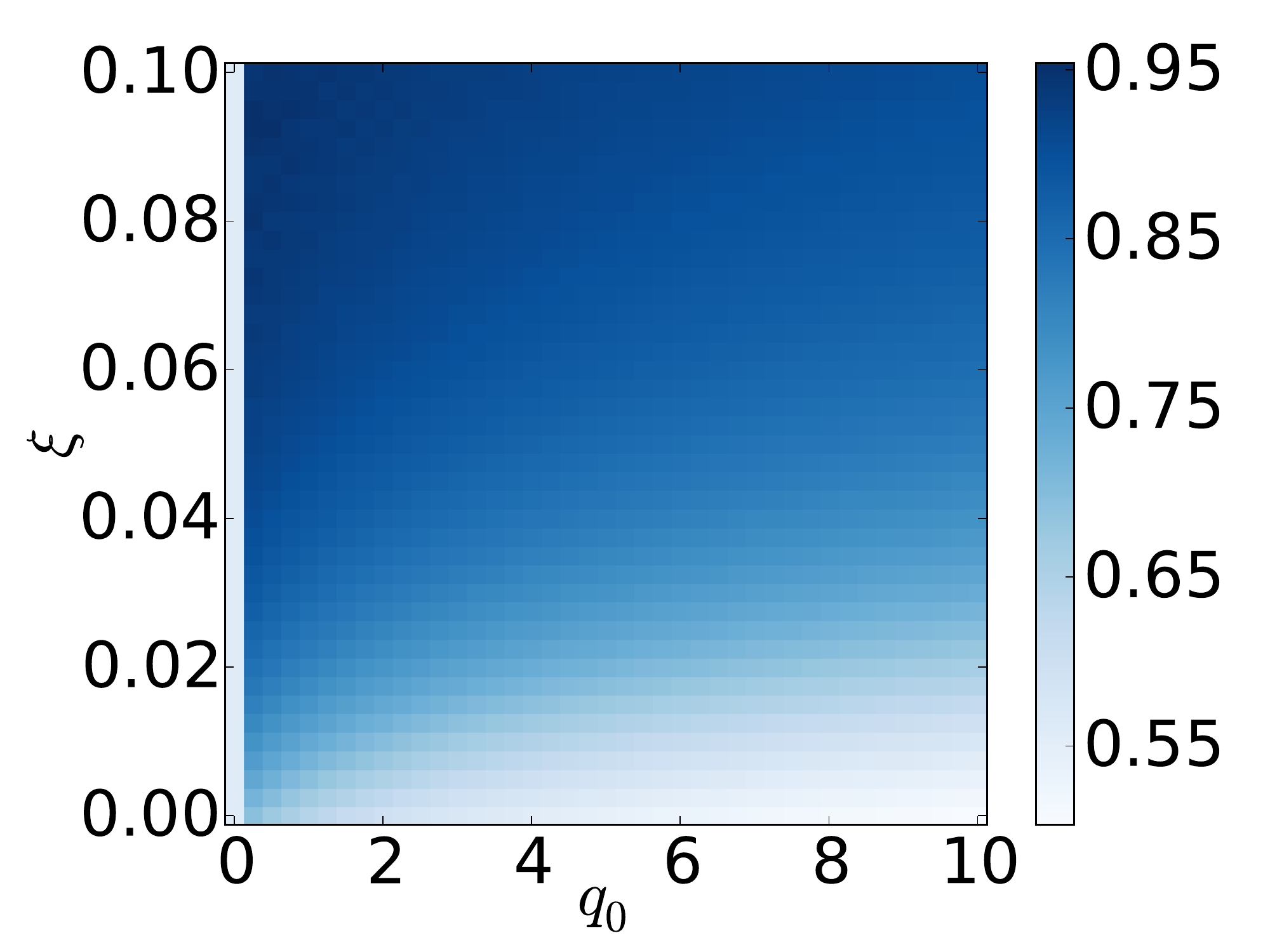}
\caption{}
\end{subfigure}

\begin{subfigure}{0.49\textwidth}
\includegraphics[width=\textwidth]{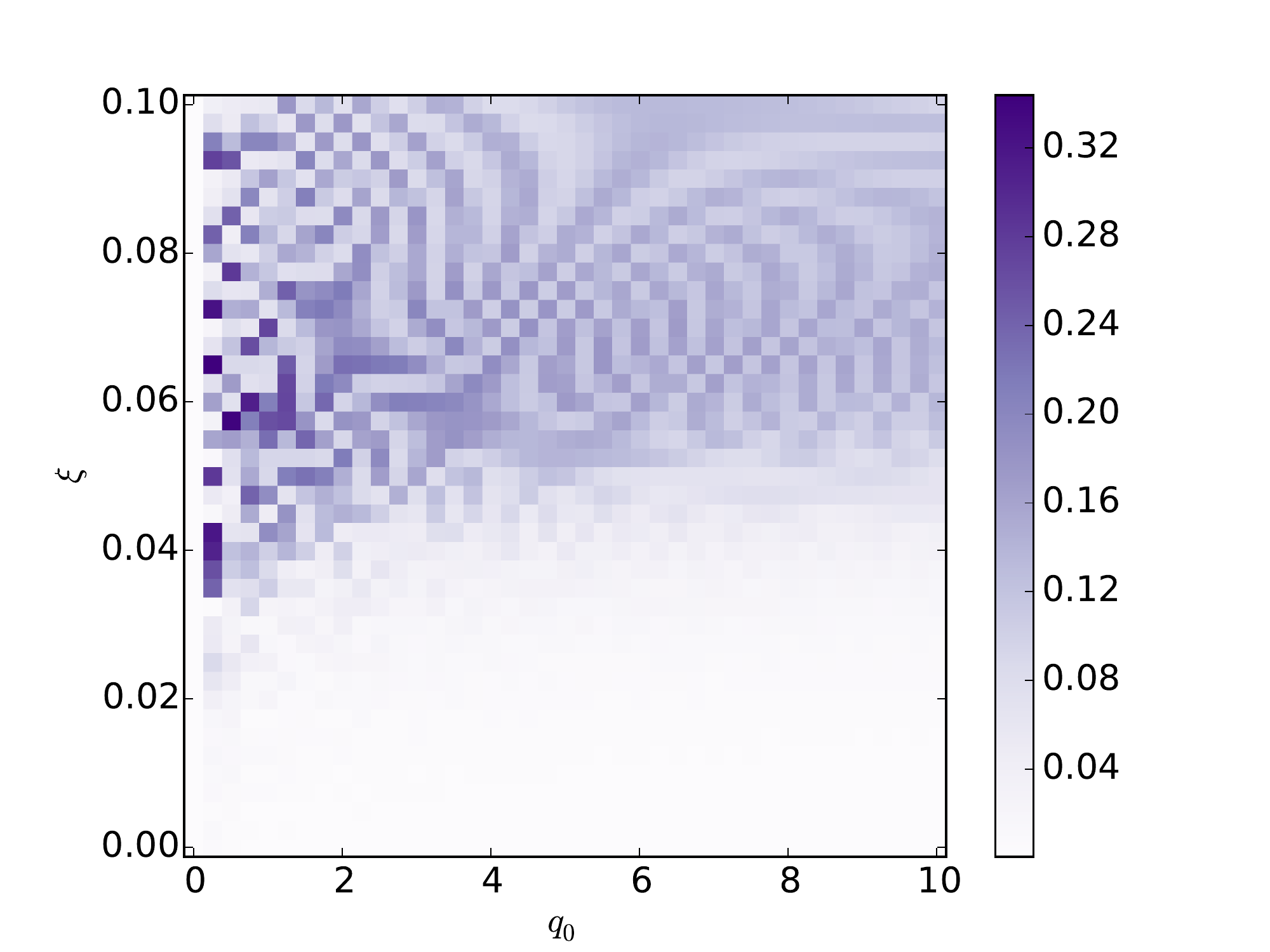}
\caption{}
\end{subfigure}
\caption{Simulations for $N=1000$ atoms, with a sweep $q(t)/\Lambda = q_0 \mathrm{e}^{-0.05\Lambda t}$, to a final time $\Lambda t = 200$. (a) Overlap with the spin singlet, (b) probability of no photons being emitted during the sweep, and (c) the product of those two, giving the efficiency of a combination of sweep and dissipative distillation.\label{largeatoms}}
\end{figure}

Since we are interested in only the final overlap with the singlet state, we can reduce our basis to those states accessible before a jump has occurred; that is, we consider only states with an exact number of pairs in the $m=\pm1$ states, $\ket{k,N-2k,k}$. Such a reduction in basis allows us to greatly increase the number of atoms we simulate. We model the back-action of the null measurement by adding a non-Hermitian term to the Hamiltonian,
\begin{equation}
\hat{H} = \frac{\Lambda-i\Gamma}{N}(\hat{S}_x^2 + \hat{S}_y^2) + q(t)\hat{N}_0 ,
\end{equation}
and calculate the probability of no photons being produced during the sweep from the jump operator expectation value,
\begin{equation}
p_s = \prod\limits_{i=0}^{t_{\mathrm{max}}/\mathrm{d}t} \left\{ 1-2\Gamma \braket{\hat{S}_x^2 + \hat{S}_y^2}_{t_i} \mathrm{d}t \right\} ,
\end{equation}
where $\braket{\hat{\mathcal{O}}}_{t_i}$ is the expectation of the operator at time $t = i\,\mathrm{d}t$. This means that not only do we vastly reduce the basis size, but we also have only to integrate the Schr\"{o}dinger equation once, rather than running a large number of trajectories.

Fig.~\ref{largeatoms} shows the overlap and post-selection efficiency for $N=1000$ atoms. We can see a smooth relationship between the sweep parameters and the post-selection efficiency. Simply, the sweep should be fast and from a small, non-zero $q_0$. Optimisation of the overlap is more complicated, though we see a wide range of parameters for which that overlap is significantly enhanced. The product of these thus gives a wide region of sweep parameters for which the total efficiency to produce a spin singlet of near perfect fidelity for 1000 atoms would be $\sim 10-20\%$. Even allowing for experimental reductions to this efficiency, such a method would produce the state frequently enough to allow for study and, potentially, use of the many-body entangled state.

For potential parameters, we consider parameters for the Tavis-Cummings model of $\set{\lambda_-,\kappa}/(2\pi) \simeq \set{300,10}$ kHz. Large $\omega/2\pi \simeq 10$ MHz would then produce the spinor collisional model necessary for the quasi-adiabatic sweep. Switching to the dissipative method can then be performed by rapidly changing $\omega \rightarrow 0$. From a microscopic perspective, this involves shifting the frequency of the $\sigma_-$-polarized laser closer to the cavity frequency. In this limit, we have an open Tavis-Cummings model, which would perfectly project the singlet state on the order of 1~ms or less. With these parameters, the scenario of Fig.~\ref{largeatoms} corresponds to a sweep that lasts 3.2~ms, and hence to a time for the total procedure of approximately 4~ms.

\section{Conclusion}

We have proposed a method to prepare the spin-singlet state of an ensemble of integer-spin atoms. Despite great interest in such a state, the production of the singlet is an open problem. Its preparation in spinor BEC experiments faces significant challenges; one potential method is that of adiabatic transformation, but this is made very difficult by the tiny energy gap at the phase transition, which greatly inflates the required timescales, as well as by the need to thoroughly minimize any residual magnetic field. 

Our methods bypass these issues by using an alternative approach based upon engineered dynamics and projective measurement in cavity QED.
Using a scheme of cavity-assisted Raman transitions to produce an effective spin-1 Tavis-Cummings model, we make use of the fact that the spin singlet is a unique dark state of the system. This means that the production of the singlet state is heralded by the \emph{absence} of photons detected in the cavity output channel. 

Using a slightly modified scheme of cavity-assisted Raman transitions enables us to emulate spinor BEC dynamics and thereby implement a quasi-adiabatic transformation to enhance the overlap of the atomic state with the singlet state prior to implementing the effective Tavis-Cummings model, thus enhancing the probabilistic distillation of the singlet state through monitoring of the dissipative output channel. We believe that this approach offers a realistic possibility of reliably producing the spin-singlet state experimentally, and that it highlights the potential of augmenting spinor interaction models with dissipative channels such as cavity modes.

\acknowledgments
The authors acknowledge the contribution of NeSI high-performance computing facilities to the results of this research. New Zealand's national facilities are provided by the New Zealand eScience Infrastructure and funded jointly by NeSI's collaborator institutions and through the Ministry of Business, Innovation and Employment's Research Infrastructure program.

\bibliographystyle{apsrev4-1}

\end{document}